\documentclass[12pt,preprint]{aastex}

\newcommand\rsun{\hbox{\,R$_\odot$}}

\newcommand\kms{ km~s$^{-1}$}

\def\kms{\ifmmode{\rm km\thinspace s^{-1}}\else km\thinspace s$^{-1}$\fi}

\shorttitle{Closure Phase Calibration Study}
\shortauthors{Zhao et al.}

\begin{document}

\title{Toward Direct Detection of Hot Jupiters with Precision Closure Phase: Calibration Studies and First Results from the CHARA Array}

\author{M.~Zhao\altaffilmark{1},
J.~D.~Monnier\altaffilmark{2},
X. Che\altaffilmark{2},
E. Pedretti\altaffilmark{3},
N. Thureau\altaffilmark{3},
G. Schaefer\altaffilmark{4},
T. ten Brummelaar\altaffilmark{4},
A. M{\'e}rand\altaffilmark{5},
S. T. Ridgway\altaffilmark{6},
H. McAlister\altaffilmark{4},
N. Turner\altaffilmark{4},
J. Sturmann\altaffilmark{4},
L. Sturmann\altaffilmark{4},
P. J. Goldfinger\altaffilmark{4},
C. Farrington\altaffilmark{4}
}

\altaffiltext{1}{ming.zhao@jpl.nasa.gov: Jet Propulsion Laboratory, California Institute of Technology,  4800 Oak Grove Dr, MS 169-327, Pasadena, CA 91109}
\altaffiltext{2}{University of Michigan, Astronomy Department, USA}
\altaffiltext{3}{SUPA, University of St. Andrews, Scotland, UK}
\altaffiltext{4}{The CHARA Array, Georgia State University}
\altaffiltext{5}{European Southern Observatory}
\altaffiltext{6}{National Optical Astronomy Observatory, NOAO, Tucson, AZ}

\begin{abstract}

Direct detection of thermal emission from nearby hot Jupiters has greatly advanced our knowledge of extrasolar planets in recent years.  
Since hot Jupiter systems can be regarded as analogs of high contrast binaries, ground-based infrared long baseline interferometers 
have the potential to resolve them and detect their thermal emission with precision closure phase - a method that is immune to the systematic errors induced by the Earth's atmosphere.
In this work, we present closure phase studies toward direct detection of nearby hot Jupiters using the CHARA interferometer array outfitted with the MIRC instrument. We carry out closure phase simulations and conduct a large number of observations for the best candidate $\upsilon$ And.
Our experiments suggest the method is feasible with highly stable and precise closure phases.  
However, we also find  much larger systematic errors than expected in the observations, most likely caused by dispersion across different wavelengths. We find that using higher spectral resolution modes (e.g., R=150) can   significantly reduce the systematics. 
By combining all calibrators in an observing run together, we are able to roughly re-calibrate the lower spectral resolution data, allowing us to obtain upper limits of the star-planet contrast ratios of $\upsilon$ And b across the $H$ band. The data also allow us to get a refined stellar radius of 1.625$\pm$0.011\rsun.
Our best upper limit corresponds to a contrast ratio of 2.1$\times10^3$:1 with 90\% confidence level at 1.52$\mu$m , suggesting that we are starting to have the capability of constraining atmospheric models of hot Jupiters with interferometry.
With recent and upcoming improvements of CHARA/MIRC, the prospect of detecting emission from hot Jupiters with closure phases is promising. 
\end{abstract}


\section{Introduction}
\label{intro}  

The discovery of a planet around a nearby star 51 Peg in 1995 opened a window into new worlds
outside the solar system \citep{Mayor1995}. Since then, more than 500 so-called exoplanets\footnote{data from The Extrasolar Planets Encyclopaedia: http://exoplanet.eu/catalog.php}
have been discovered, revolutionizing our knowledge of their nature and origin. Among
those discoveries, about 24 planets had their thermal emission directly detected by photometric and/or spectroscopic measurements from space or ground, most of which are known as ``hot Jupiters" or ``hot Neptunes" \citep[e.g.,][etc.]{Beerer2010, Machalek2010, Nymeyer2010}.
So far, direct detection of thermal emission and characterization of planetary atmospheres is generally only possible for these hot planets as they are very close to their host stars ($<$0.1AU), 
and are thus heated enough to have temperatures above 1000K, providing as high as $10^{-3}$ of their host stars'  flux in the infrared (e.g., $J$, $H$, $K$, and mid-infrared).

The atmospheres of hot Jupiters have many interesting properties. 
For instance, transmission spectra at primary eclipses have shown the presence of sodium, water and methane \citep[e.g., ][]{Redfield2008, Swain2009}, while thermal emission at secondary eclipses has shown the presence of water, carbon dioxide, and carbon monoxide \citep[e.g.,][]{Knutson2007b, Barman2008, Swain2008}. Due to their close-in orbits, hot Jupiters are tidally locked to their host stars, leading to  a constant day side that experiences intense stellar irradiation and a cold night side that remains in perpetual shadow \citep{Guillot1996}. The temperature difference between the day and night sides thus induces atmospheric circulation and strong zonal winds to redistribute the heat \citep{Knutson2007}.  
Studies have also suggested the existence of a hot stratosphere on the dayside of some hot Jupiters (e.g., HD 209458b and HD 149026b), which inverts the temperature profile of higher atmospheric layers, and  flips water absorptions into emissions \citep{Knutson2008, Burrows2008, Madhusudhan2010}. However, some other planets such as HD 189733b seem to lack such thermal inversion \citep{Charbonneau2008, Grillmair2008}, implying fundamental atmospheric differences between these planets and leading to sub-classification of hot Jupiters \citep{Burrows2008}. Studies of hot Jupiters' atmospheres will not only reveal their composition, structure, and dynamics, but will also  shed light on our understanding of the planet formation processes. More importantly, characterizing hot Jupiters allows us to pave the path
toward characterizations of Super-Earth planets and eventually, Earth-like planets.

Among the $\sim22$ detected hot Jupiters, however, a majority of them are transiting planets whose orbits are aligned with the line of sight from
the Earth, and only 2 of them are non-transiting planets (i.e., $\upsilon$ And b and HD 179949b, \citet{Harrington2006, Cowan2007}). 
The lack of studies of non-transiting hot Jupiters leaves a great opportunity for long baseline optical/infrared interferometry, in that long baseline interferometers can see hot Jupiter systems as extremely high-contrast binaries, and thus can directly determine their  orbital elements and provide accurate mass estimates. 
Interferometric measurements can also provide absolute planet/star flux ratios (or star/planet contrasts) of non-transiting planets which cannot be disentangled by combined-light techniques (i.e., through transits). In fact, interferometric measurements also provide an independent way of characterizing transiting planets in addition to the combined light technique. 
 Since the bulk of energy from hot Jupiters emerges from the near-IR between 1-3 $\mu$m \citep{Burrows2008}, interferometry can provide a better understanding of their spectra and global energy budget with measurements at near-IR bands. 

However, detecting the weak emission from a hot Jupiter from the ground is a difficult task. To date, only a few hot Jupiters' thermal emission has been detected from the ground, while the rest were all achieved from space. 
To reach this goal, we  require very stable and high precision measurements, and most importantly, require a method that can eliminate or calibrate the effects of Earth's atmospheric turbulence.  One possible approach is to use high precision and high resolution closure phase measurements obtained with ground-based interferometers.  Closure phase is measured by combining the phases of three baselines in a closed triangle. 
It is immune to any phase shifts induced by the atmospheric turbulence, including the differential chromatic dispersion that affects the differential phase. The major biases or systematic errors of closure phase come from non-closed triangles introduced in the  measurement process, and in principle, can be precisely calibrated. 
Therefore, it is a good observable for stable and precise measurements.  
More descriptions of the closure phase technique can be found in \citet{Monnier2003, Monnier2007}.

Studies have already been carried out to explore the possibilities of using closure phases for exoplanet detection \citep[e.g.,][etc.]{Joergens2004, Zhao2008b}. Particularly, \citet{Chelli2009} studied in detail the characteristics of closure phases and the corresponding SNR when the primary star of a binary is getting resolved and approaching the visibility null (i.e, "Phase Closure Nulling"). Using this method, \citet{Duvert2010} detected the faint close companion of HD 59717 at a contrast of $\sim100:1$ using VLTI/AMBER. Recently, \citet{Absil2010} also excluded the presence of a brown-dwarf companion in the innermost region of the $\beta$ Pic planetary system at a upper contrast limit of $\sim200:1$ using closure phases obtained by VLTI/AMBER. 

In this paper, we report our closure phase studies toward the ultimate goal of directly detecting emission from hot Jupiters using CHARA/MIRC.
The paper is organized as follows. We first briefly introduce our candidate $\upsilon$ And b and  our observations in \S\ref{obs}. In \S\ref{simu} we simulate the closure phase signals and the required precision  for the candidate. We  then discuss the calibration issues we encountered in test observations and present our solutions. Based on our calibrations, we present a preliminary upper limit for $\upsilon$ And b in \S\ref{upsand}. Finally, we conclude our studies and give future prospects in \S\ref{conclusion}.

\section{Candidates and observations}
\label{obs}

\subsection{Candidates}
Among all the known non-transiting hot Jupiters, several of them are within the sensitivity limit of CHARA/MIRC. $\upsilon$ And b is currently the most favorable candidate due to its relatively high temperature and the high brightness of its host star. Therefore, in this study we focus only on this best candidate.

 $\upsilon$ And is an F8V star located 13.5 pc away from the Sun. \citet{Butler1997} first discovered its hot Jupiter $\upsilon$ And b in 1997, which has a period of 4.6 days and is orbiting at 0.06 AU. The follow-up observations of \citet{Butler1999} found two more companions in the system, $\upsilon$ And c and $\upsilon$ And d. $\upsilon$ And c is orbiting at 0.83 AU from the host star with a period of 241 days, while $\upsilon$ And d is orbiting at 2.5 AU with a period of 1267 days \citep{Butler1999}. Most recently, \citet{Curiel2011} found a fourth planet $\upsilon$ And e with a period of 3848.9 days at 5.25 AU.  The system is non-coplanar.  The inclination of the closest planet $\upsilon$ And b is found to be likely $\geq28^o$ \citep{Crossfield2010}, while the second and the third planets ($\upsilon$ And c \& d) have a relative inclination of 15$^o$-20$^o$.
In 2006, \citet{Harrington2006} directly detected thermal emission from the hot Jupiter $\upsilon$ And b  using Spitzer MIPS at 24$\mu m$, in which they
detected the relative day-night flux variations of the planet over five epochs of the whole 4.6-day orbital period, and provided a lower limit to the planet/star flux ratio. Later, \citet{Crossfield2010} further refined the flux variation curve with more MIPS 24$\mu m$ data, and found an unusually large phase shift of the flux maximum ($\sim80^o$). Atmospheric models have also been  applied to interpret these Spitzer data. However, as \citet{Burrows2008} pointed out, due to the lack of absolute flux level and information in other wavelengths, there are too many degrees of freedom to draw strong conclusions about the planetary and atmospheric properties. 
Thus, detection of its absolute planet/star flux ratios and at other wavelengths such as near-IR are  required to break model degeneracies. Furthermore, high signal-to-noise detections will even allow us to obtain the absolute phase curve of the planet \citep{Barman2005}, providing constraints to the circulation and heat redistribution patterns of its atmosphere.

\subsection{Observations}
Closure phase measurements require at least three telescopes to be combined in a closed triangle. Currently, the MIRC \citep{Monnier2004} and CLIMB \citep{Sturmann2010} instruments at the CHARA array, and the AMBER \citep{Petrov2007} and PIONIER\footnote{http://www-laog.obs.ujf-grenoble.fr/twiki/bin/view/Ipag/Projets/Pionier/WebHome} instruments at the VLTI have this capability in the near-IR. In this study, we employ CHARA/MIRC for our measurements. 

The
 CHARA array, located on Mt. Wilson, consists of six 1-meter telescopes \citep{Brummelaar2005}.  
 The array is arranged in a Y-shaped 
 configuration to provide good position angle coverage.  The six telescopes form 15 baselines ranging from 34m to 331m,  making CHARA the longest-baseline optical/IR 
 interferometer array in the world and providing a resolution of $\sim$0.5 mas in the $H$ band. 
The Michigan Infra-Red Combiner (MIRC) is an imaging combiner that 
  currently combines  4 CHARA telescopes, providing 6 visibilities, 4 closure phases and 4 triple amplitudes simultaneously. MIRC  works at both $H$ and $K$ bands, and has three spectral modes:  R=40 (with prism), 150, \& 500 (with grism).
  The lowest resolution mode (R=40) disperses light into 8 spectral channels on the detector, while the R=150 and R=500 modes disperse light  into 24 channels and 80 channels respectively \citep[see][for details]{Monnier2004, Monnier2006}.  
The compact design of MIRC allows for stable calibration and precise closure phase measurements,  thus it is best suited for the purpose of this study.

We conducted observations of $\upsilon$ And on  23 nights  from 2006 to 2010 with a total integration time of 8.2 hours,  following the standard observing procedures \citep{Monnier2007sci, Zhao2009}. 
The observation log is listed in Table \ref{obslog}.
 Several combinations of four CHARA  telescopes are used in the observations, while in most  cases we adopt the combination S1-E1-W1-W2 and S2-E2-W1-W2 for good baseline coverage. The observations were mostly conducted in the $H$ band ($\lambda$=1.5 - 1.8 $\mu m$) with the lowest resolution mode of R=40, except for two nights in 2010 when we employed the higher resolution mode of R=150. Each observation of the target was bracketed with calibrators for visibility and closure phase calibration. For the purpose of bias subtraction and flux calibration, each set of fringe data is bracketed with measurements of background (i.e., data taken with all beams closed),  shutter sequences (i.e., data taken with only one beam open at a time to estimate the amount of light coming from each beam), and foreground (i.e., data taken with all beams open but without fringes) \citep{Pedretti2009}. Each object is observed for multiple sets. During the period of taking fringe data, a group-delay fringe tracker is used to track  the fringes  \citep{Thrueau2006}. In order to track the flux coupled into each beam in ``real time" to improve the visibility measurements, we  used spinning choppers to temporally modulate the light going into each fiber simultaneously with fringe measurements. In 2009, photometric channels were commissioned for MIRC \citep{Che2010}, and the choppers were made obsolete because of the much better real-time flux calibration provided by the photometric channels. The data reduction process also follows the pipeline outlined in \citep{Monnier2007sci}, where the closure phases are extracted from the phase term of the complex triple amplitudes after the subtraction of background, foreground, and correction of fiber coupling efficiencies. 



\section{Closure phase simulations and calibration studies}
\label{simu}

\subsection{Simulations}

Because exoplanet-host stars and their  hot Jupiters are similar to high contrast close binaries, we simulate the closure phase signals using binary models  for $\upsilon$ And b. 
We choose to use the longest telescope triangle of the CHARA array (i.e., S1-E1-W1) in our study to obtain the highest closure phase levels \citep{Chelli2009}. 
The latest orbital properties of $\upsilon$ And b and the size of its host star are listed in Table \ref{tab2}. The Infrared planet/star flux ratios are adopted from the models of \citet{Sudarsky2003}. 

\begin{figure}[H]
\begin{center}
\includegraphics[width=2.7in, angle=90]{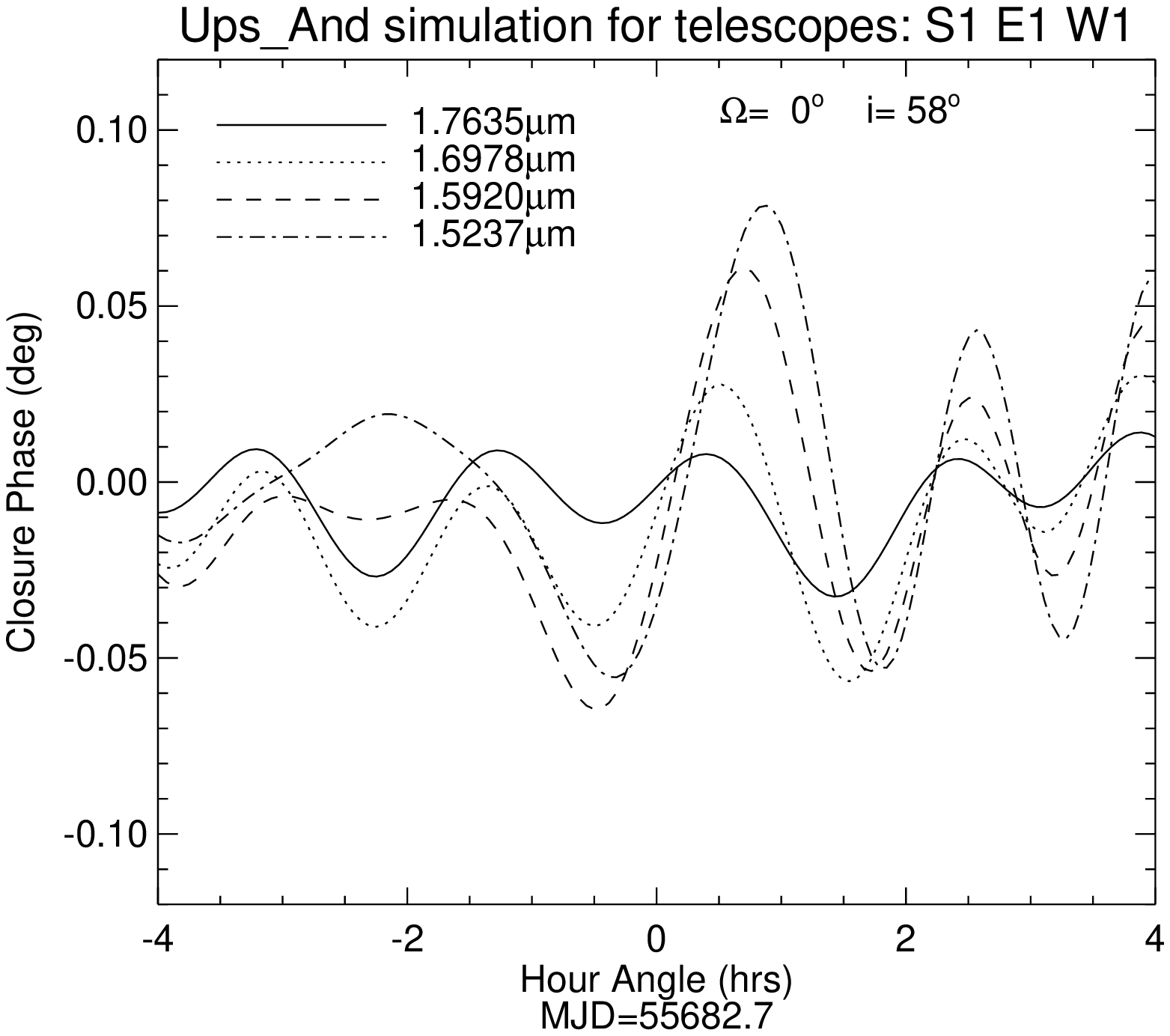}
\includegraphics[width=2.7in, angle=90]{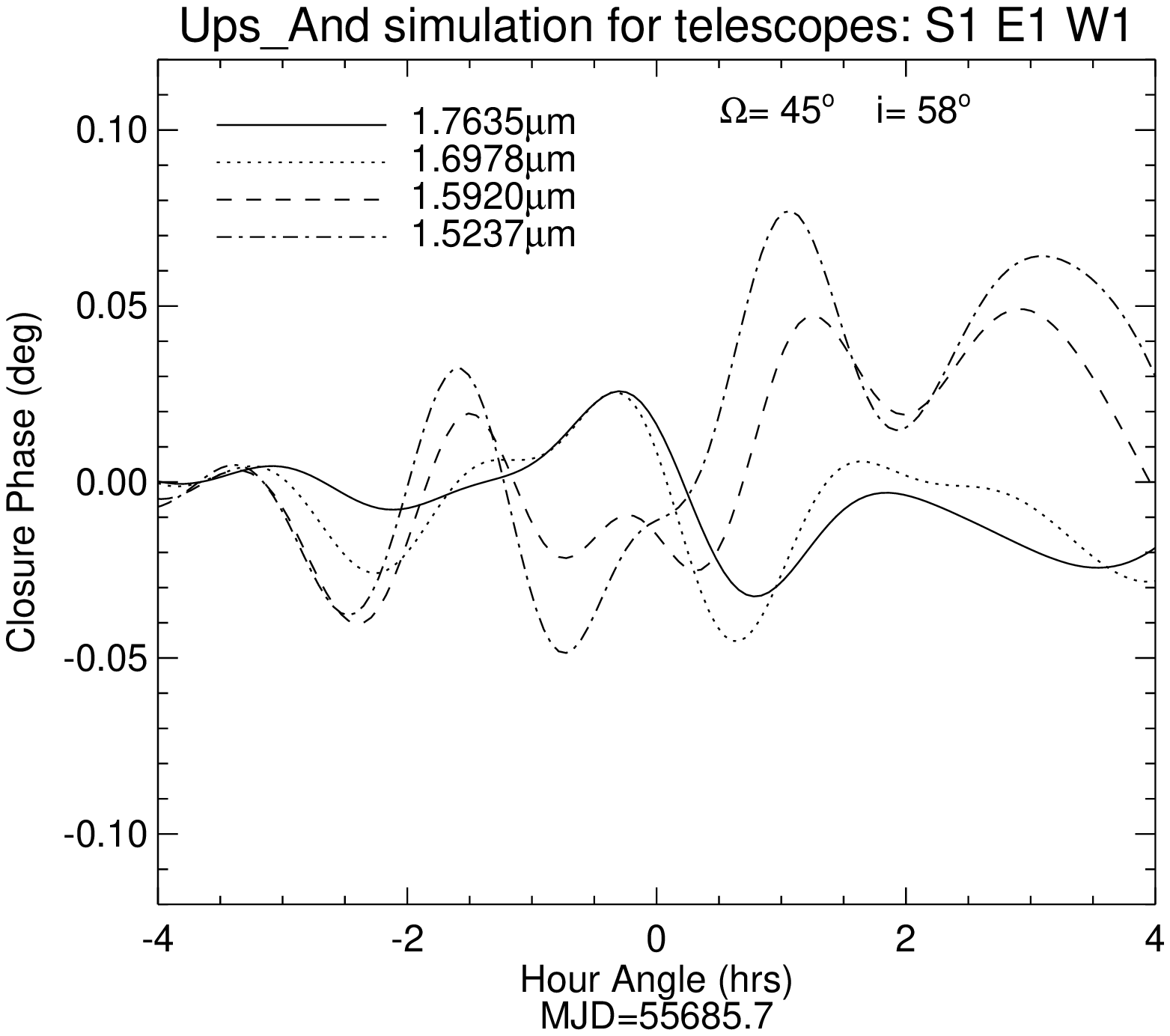} 
\includegraphics[width=2.7in, angle=90]{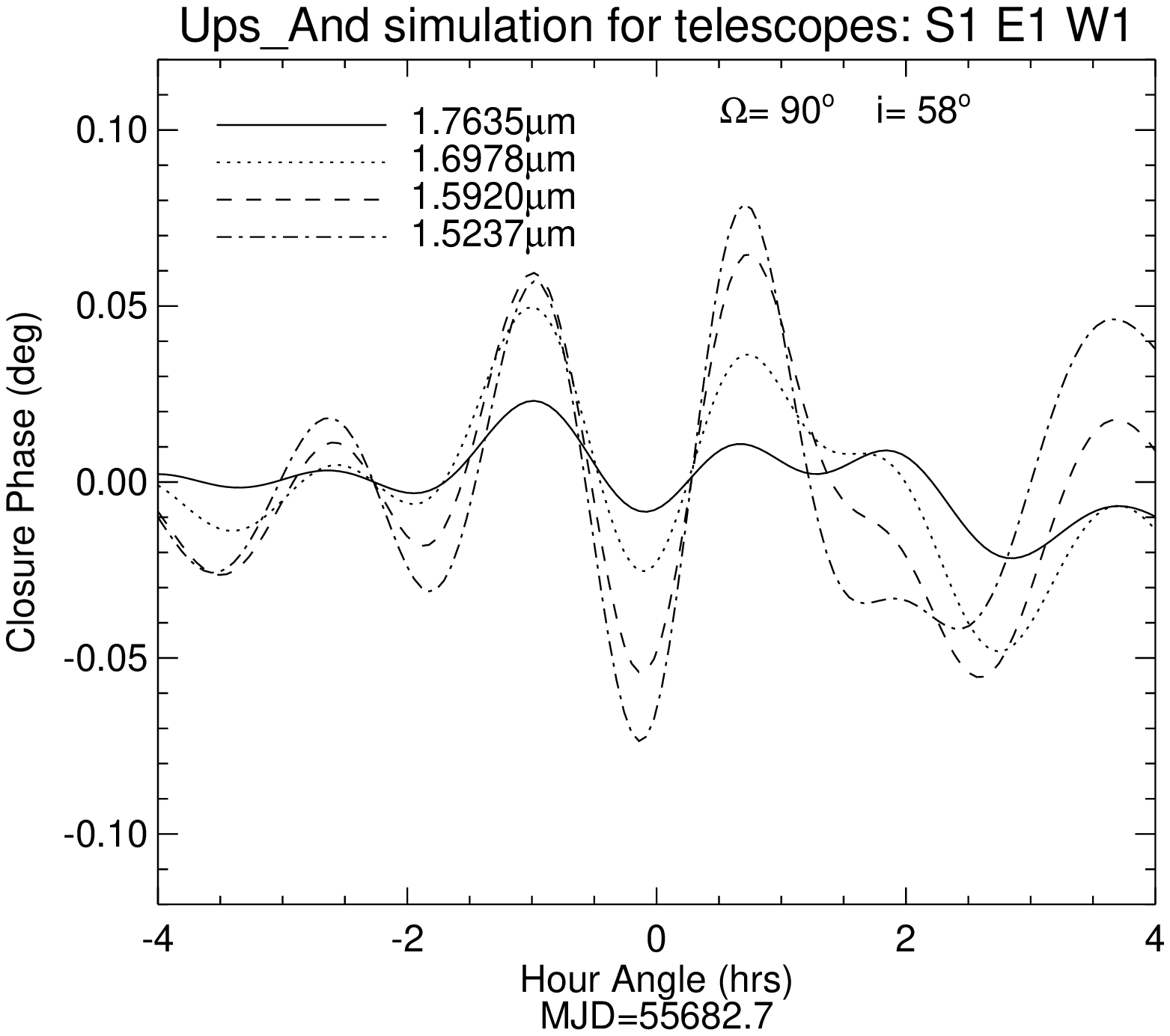} 
\includegraphics[width=2.7in, angle=90]{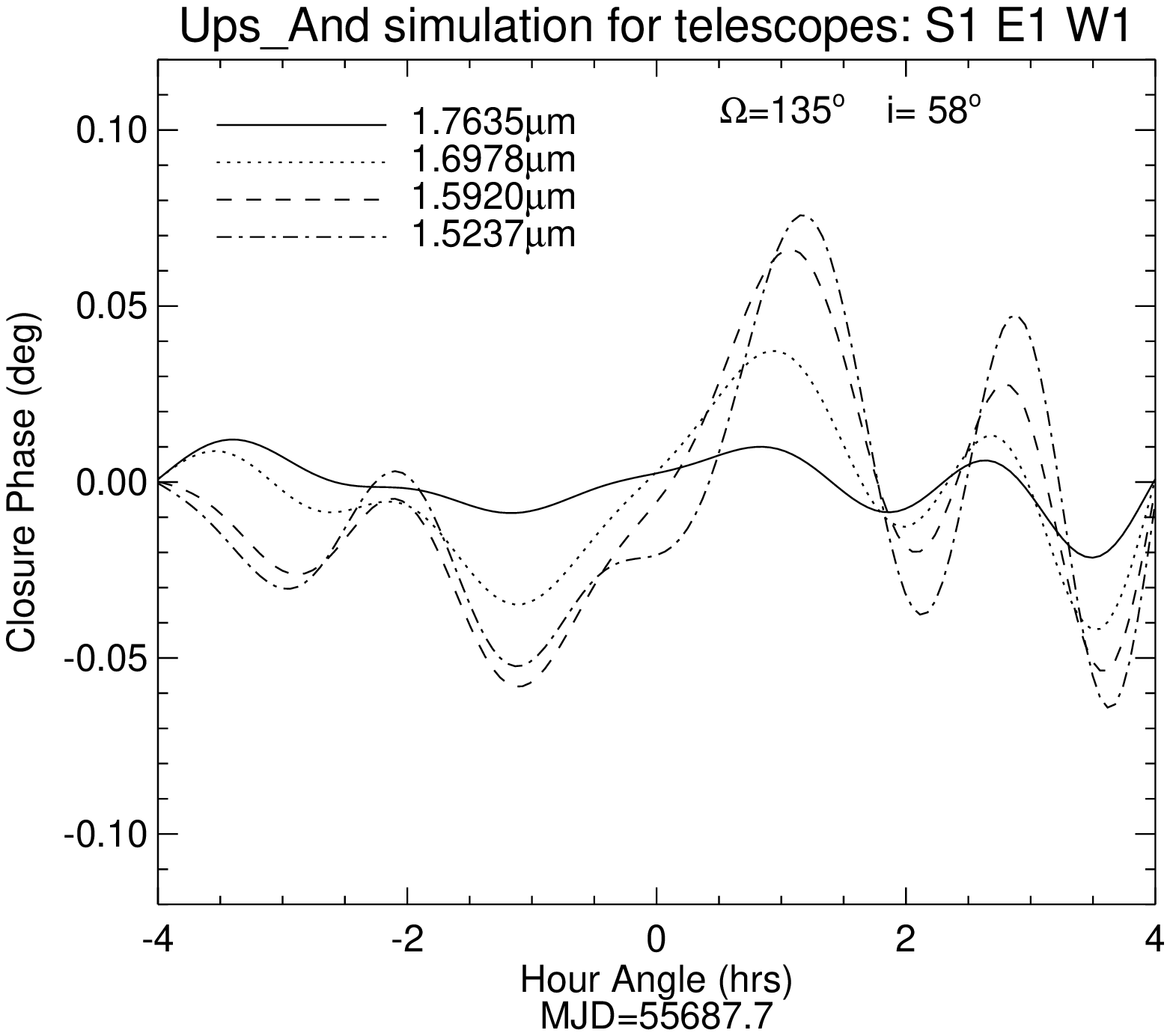} 
\caption{Closure phase simulation of $\upsilon$ And b, using the largest triangle of CHARA, S1-E1-W1, and corresponding to an angular resolution of 0.5 mas in the $H$ band. Four out of eight  wavelength channels of MIRC  in the $H$ band are shown. The planet/star flux ratios of the four channels (1.52$\mu m$, 1.59$\mu m$, 1.70$\mu m$, 1.76$\mu m$) are 0.27$\times10^{-4}$, 0.47$\times10^{-4}$, 0.6$\times10^{-4}$, and 0.35$\times10^{-4}$, respectively \citep{Sudarsky2003}. The orbital parameters used in the simulation are listed in Table \ref{tab2}. The four panels assume $\Omega=0^o, 45^o, 90^o$ and $135^o$ respectively. The dates of the simulation are chosen arbitrarily. Thanks to the even baseline layout of CHARA, the signal level of the closure phase 
is insensitive to the values of $\Omega$, although the maximum signal may occur at different times.
} 
\label{cp_figa}
\end{center}
\end{figure}

Figure \ref{cp_figa} shows the closure phase simulations for $\upsilon$ And b at four wavelength channels. Since the value of the ascending node ($\Omega$) is unknown for $\upsilon$ And b, we assume 4 different values in our simulations: $ \Omega=0^o, 45^o, 90^o, 135^o$. The inclination is fixed to the most probable values assuming that  the planet is coplanar with the star's rotation (see Table \ref{tab2}). 
Figure \ref{cp_figa} shows that the closure phase signal varies rapidly as the baseline projection varies during the Earth's rotation. Shorter wavelength channels generally  have higher signal amplitudes and peak at $\sim\pm0.08^o$, due to the fact that the host star is more resolved at those wavelengths.  
The overall closure phase level of $\upsilon$ And b is insensitive to the values of $\Omega$ in the simulation because of CHARA's even baseline layout (i.e., the Y-shape), which covers all position angles more or less equally well. We note here that because the visibility of  the star $\upsilon$ And only goes to null at certain wavelength channels, the closure phase signal in the simulation is very sensitive to the value of adopted stellar diameter. 
Figure \ref{cp_figa} also indicates that in order to detect the closure phase signal from $\upsilon$ And b, the required 1-$\sigma$ precision has to be better than roughly $\pm$0.05$^o$.

\subsection{Calibration studies }
\label{cp_result}

\begin{figure}[t]
\begin{center}
 \includegraphics[width=3.8in, angle=90]{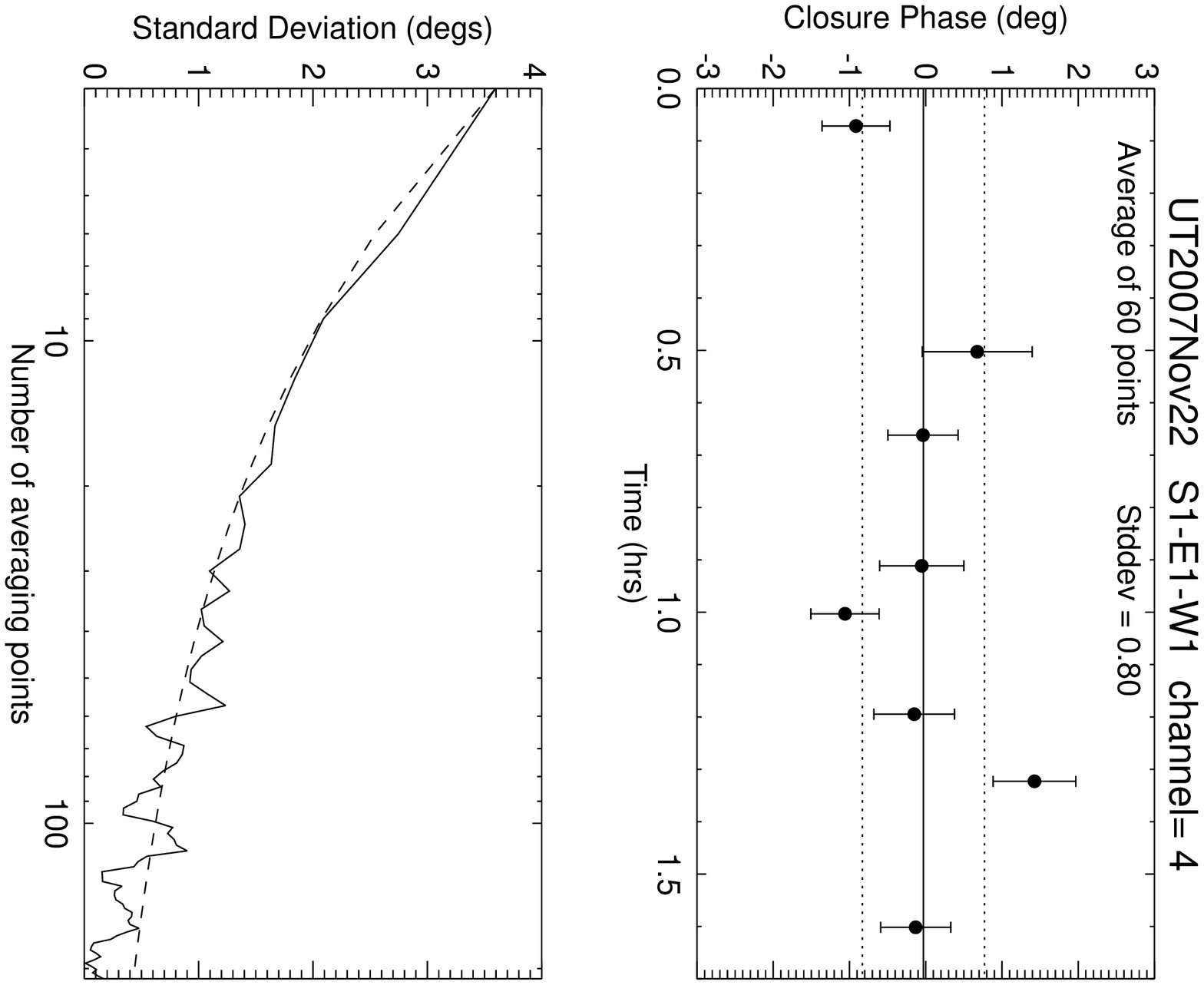}
 \includegraphics[width=3.8in, angle=90]{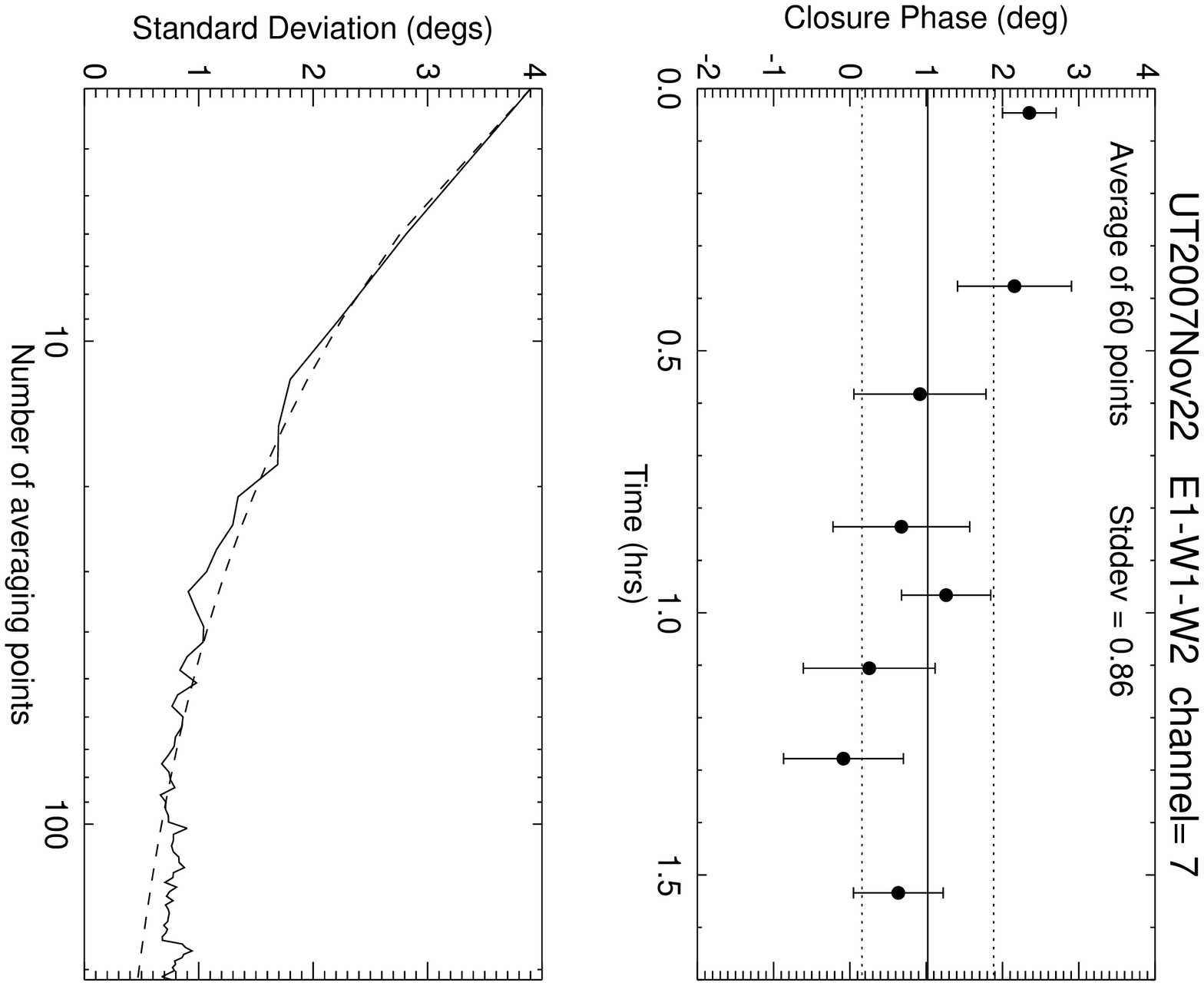}
 \caption{Test observation of $\upsilon$ And using the outer array (S1-E1-W1-W2) of CHARA from 2007Nov22. The top panels show the 60-point averaged data, corresponding to 320 sec of integration time. The solid lines in the top panels indicate the average closure phase levels, while the dashed lines indicate 1-$\sigma$ deviation from the average. 
The bottom panels compare the averaged closure phases with the standard $\sqrt{N}$ law for normally distributed errors, i.e., without systematics.  The bottom right panel shows the errors of that channel cannot be averaged down beyond 60 averaging points due to the systematic drift.
 }
   \label{cp_fig5}
\end{center}
\end{figure}

To compare the precision of our measurements with the simulations, we first examine the quality and stability of our data. 
The left panels of Figure \ref{cp_fig5} show a good night of closure phase measurements for the middle wavelength channel of MIRC, obtained with the largest triangle of CHARA (S1-E1-W1). The closure phases are stable over the 1.5 hrs of observation and the error averages down roughly as $\sqrt{N}$ (the bottom left panel), suggesting the measurements are immune from systematic errors like changes in the seeing. The nominal measurement  error is 0.3$^o$ when averaging the whole 1.5 hrs of observation together. The performance at a similar channel with a shorter triangle of CHARA (E2-W1-W2) is demonstrated to be about 3 times better in \citet{Zhao2008c}. 

\begin{figure}[t]
\begin{center}
   \includegraphics[width=5in, angle=0]{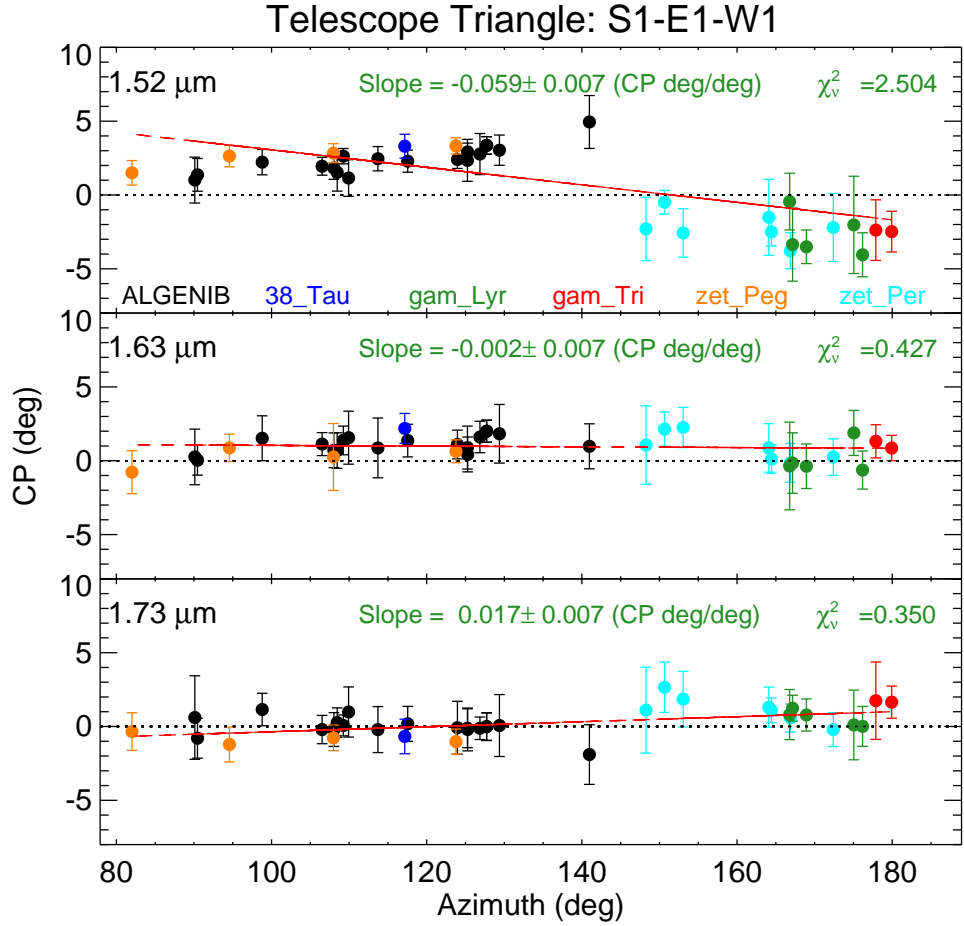}
 \caption[Closure phase vs.  Azimuth for four calibrators in 2008 August.]{Closure phase vs. Azimuth for six calibrators in 2008 August. The first, middle and the last wavelength channels of MIRC are shown from top to bottom. The data were taken with telescope S1-E1-W1. The red line shows the  linear fit of closure phase as a function of  azimuth. Different colors indicate different targets. The slope of the linear fit and the reduced $\chi^2$ are also labeled in each panel. }
\label{cpfit_az}
\end{center}
\end{figure}

\begin{figure}[t]
\begin{center}
   \includegraphics[width=5in, angle=0]{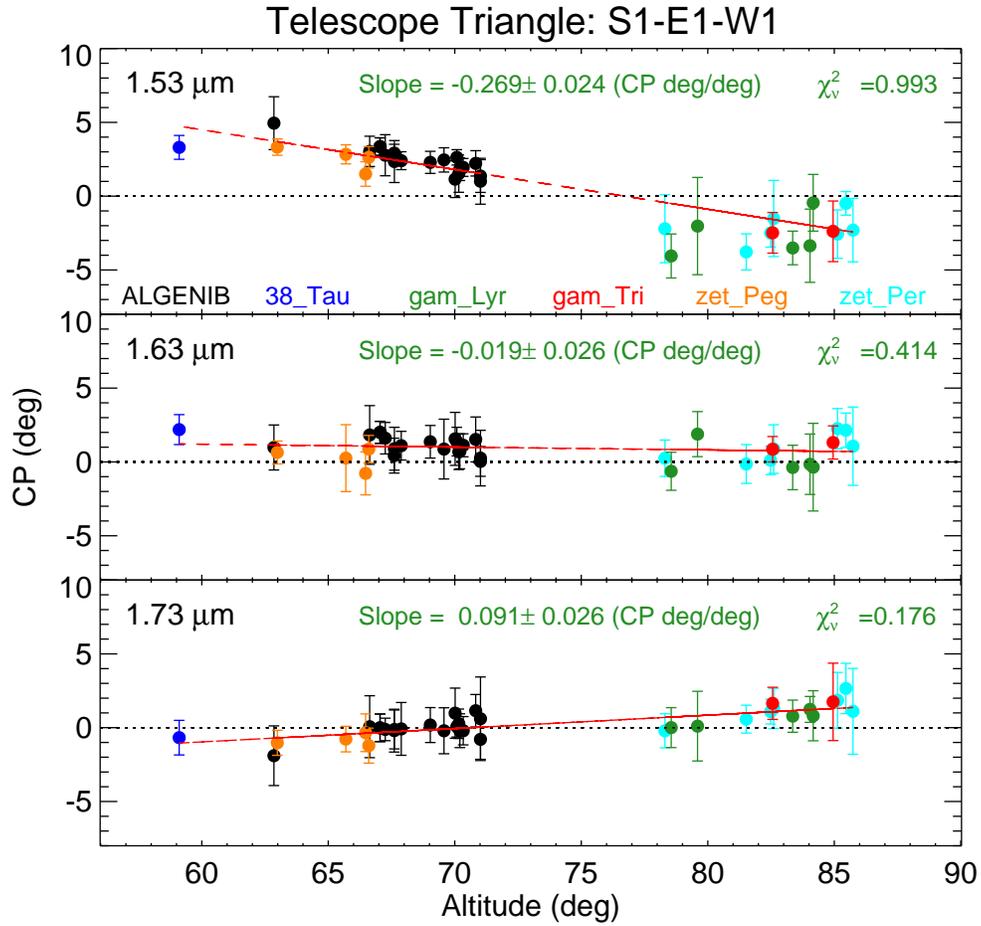}
 \caption[Closure phase vs.  Altitude for four calibrators in 2008 August.]{Closure phase vs.  Altitude for six calibrators in 2008 August. The red line shows the  linear fit of closure phase as a function of  altitude. Different colors indicate different targets. The parameters of the fits are also shown in each panel. The best-fits are good and within the scatter and error bars of the data.}
\label{cpfit_alt}
\end{center}
\end{figure}

Although these measurements are promising and stable, we have also encountered large unexpected systematic errors in other wavelength channels. As indicated in the right panels of Figure \ref{cp_fig5},  the closure phases show a systematic drift and the associated errors cannot be averaged down as $\sqrt{N}$.
In fact, the closure phases not only change with time, but also vary as a function of wavelength.
More interestingly, the closure phase drifts are highly correlated for all calibrators within a whole observing run  of many nights. 
Figure \ref{cpfit_az} plots closure phases of six calibrators obtained in 6 nights in 2008 August (Algenib, 38 Tau, $\gamma$ Lyr, $\gamma$ Tri, $\zeta$ Peg, $\zeta$ Per) versus azimuth angles. 
There is a clear trend of closure phase change as a function of azimuth in the figure. 
 Figure \ref{cpfit_alt} shows the similar correlation of closure phase with altitude. The closure phase drift reaches about 8 degrees between the top and the bottom panel. 
In addition,  we can also see an obvious slope change centered at the middle wavelength channels in both Figures \ref{cpfit_az} \& \ref{cpfit_alt}. Although only 3 wavelength channels are shown here, the slope actually  changes gradually from the first to the last channels of MIRC, and similar effects are also seen in other observing runs.  
 Although the data shown in these figures were obtained in 6 nights, the correlations with altitude and azimuth are strong and consistent,  suggesting that the major cause of the closure phase drifts may stem from the changing positions of the targets on the sky. 

Possible causes of the closure phase drifts may include: 1. polarization effects caused by the non-identical beam trains of CHARA; and 2. extra dispersions in the delaylines that are not compensated in vacuum, which can contaminate closure phases across wavelength channels and change with target position. To investigate the individual causes and find out the best solution, we carried out a series of experiments with: 1. using a linear polarizer to reduce the effect of polarization; 2. using a linear polarizer and a 40$\mu m$ slit to reduce the effect of polarization and partially reduce dispersion; and 3. using a grism of R$\sim$150 to reduce dispersion only.  We determine the slopes of the closure phase change as a function of only altitude for simplicity.
Since the slopes of the closure phases also drift across wavelength channels (see Figure \ref{cpfit_alt}), we use the magnitude of the slope change, i.e., {(slope of channel 8 $-$ slope of channel 1)}, to characterize the closure phase changes. The magnitudes of slope change are averaged over four closure phase triangles of each observing run, and the errors are determined from the scatter of the four triangles. 
The results of the experiments are shown and compared in Table \ref{tab3}. 

Table \ref{tab3} shows that the original observations have the largest magnitude of closure phase change. Observations with polarizer and polarizer + 40$\mu m$ slit have slightly smaller magnitudes of  closure phase change. However, we have also seen some nights without polarizers have similar or even lower drifts than those with the polarizer, indicating the effect of using a polarizer is small  and the major cause of the drifts may not be polarization effects. 
When using the grism of R=150, however, the correlated closure phase drifts and slope change become nearly zero. Although we only had a very small amount of data for the experiments, this preliminary result indicates that the extra air dispersion from the delaylines is most likely the major cause of the closure phase drifts, which contaminates closure phases with non-closed triangles from other wavelengths, especially at the edges of the bandpass. The dispersion effect also explains the slope change centered at the middle wavelength channels seen in Figures \ref{cpfit_az} \& \ref{cpfit_alt}. 

\begin{figure}[t]
\begin{center}
   \includegraphics[width=5in, angle=0]{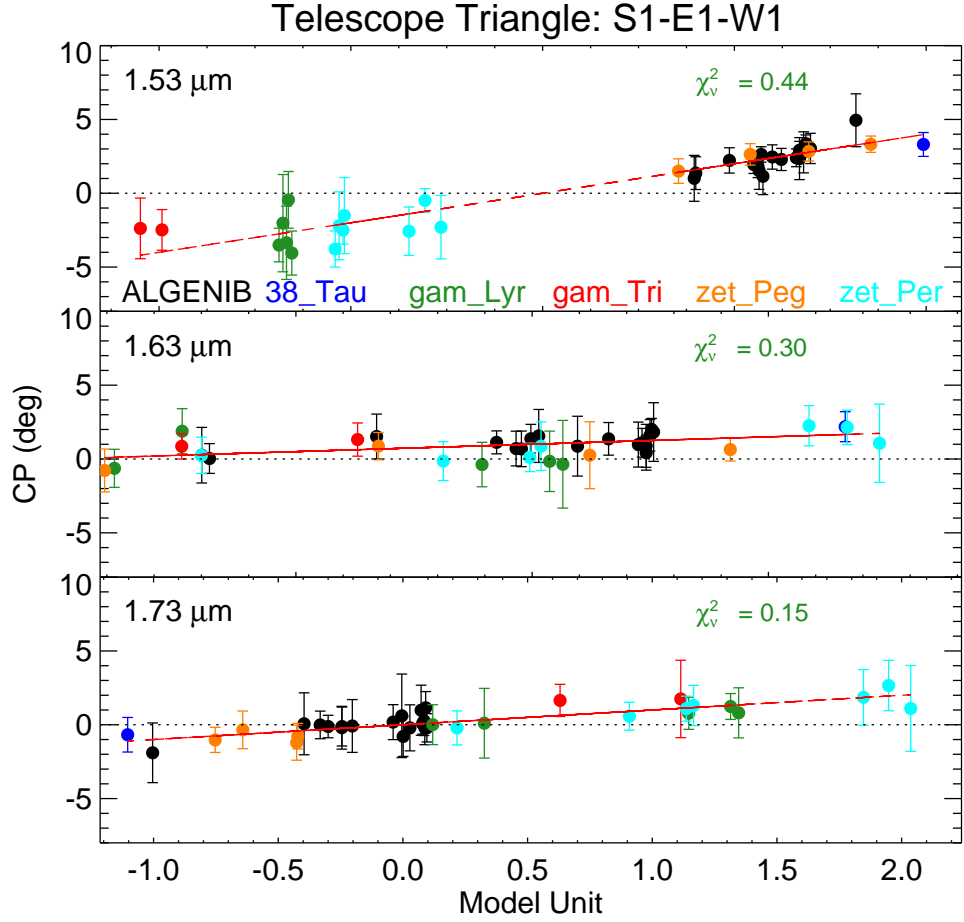}
\caption[Closure phase as a function of azimuth and altitude using quadratic fit]{Closure phase as a function of azimuth and altitude using quadratic fit.  The arrangement and notations are similar to those of Figures \ref{cpfit_az} and \ref{cpfit_alt}, except that the red line is a fit to the quadratic plane function: $a_0 + a_1\cdot Az  + a_2\cdot Az^2 + a_3\cdot Az \cdot Alt + a_4\cdot Alt + a_5\cdot Alt^2$.  The reduced $\chi^2$s shown in each panel indicate significant improvement than those of Figures \ref{cpfit_az} and \ref{cpfit_alt}.}
\label{cpfit_qua}
\end{center}
\end{figure}

Since the closure phase changes shown in Figures \ref{cpfit_az} \& \ref{cpfit_alt} are highly correlated for all nights within an observing run with consistent system and optical settings. We can use all calibrators from a run to look for a solution to calibrate the drifts. We have experimented with several function forms to characterize the slopes, including linear function, linear surface, quadratic surface, etc. A linear surface fit can estimate the closure phase drifts well, while a quadratic surface fit works the best.  
Figure \ref{cpfit_qua} shows an example of our best approach of fitting the closure phases with quadratic surface functions of both target altitude and azimuth.  The quadratic surface function characterizes the drifts very well for nearly all wavelength channels, as indicated by the improved $\chi^2$ values in the plot. It therefore can be employed as a new empirical model to calibrate our data within the same observing run. 

 
There is also a caveat, however, that this calibration scheme requires a wide span of calibrator positions on the sky for a good coverage of altitude and azimuth to bracket the target, so that the quadratic fit can reliably estimate the closure phase change. However, most of our observations have a limited number of calibrators and calibrator visits, and thus do not have a wide range of position coverage. Therefore, we adopt the results from a linear surface fit for those cases instead in order to avoid unreliable or erroneous quadratic extrapolation, with a trade-off of slightly less accurate closure phase calibration. Further validations of these schemes and their robustness are required in a future work.

\begin{figure}[t]
\begin{center}
   \includegraphics[width=5in, angle=90]{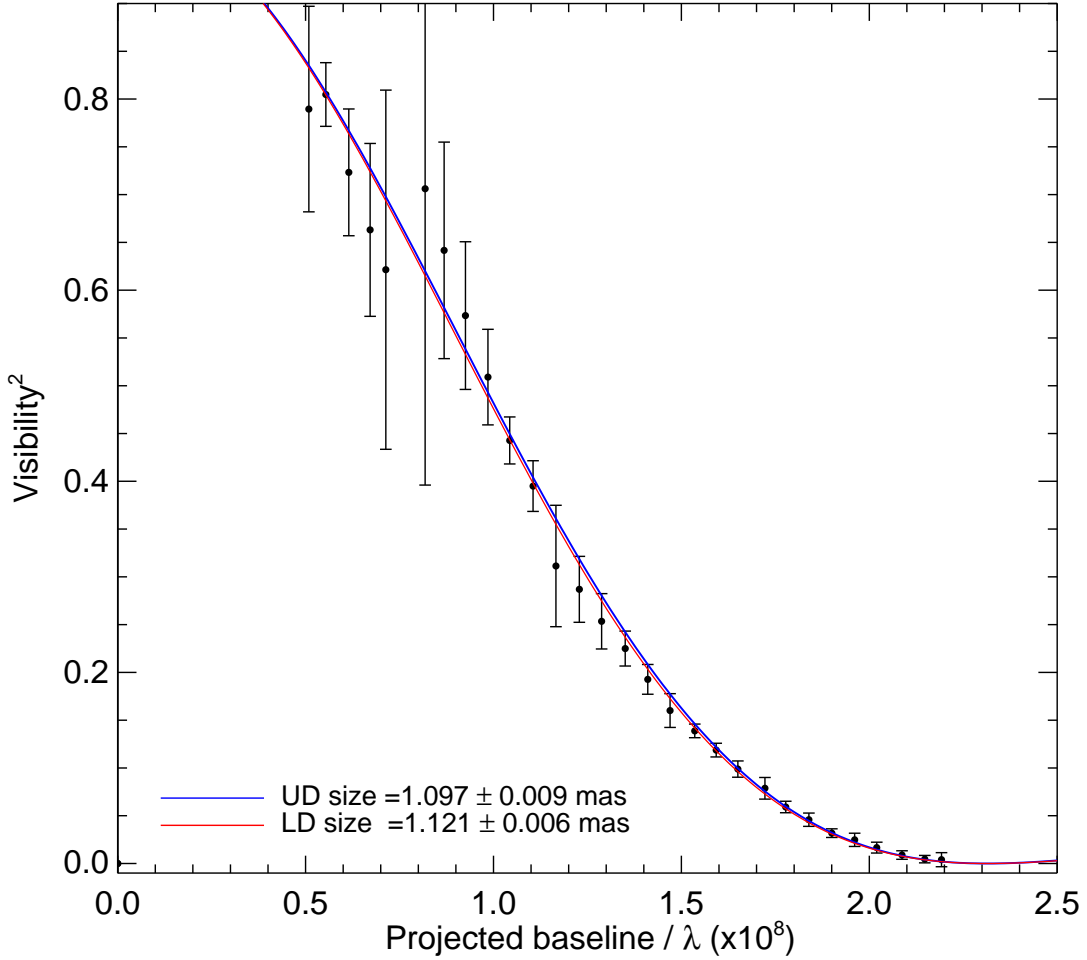}
\caption{The best-fit $V^2$ of $\upsilon$ And with a uniform disk (UD) model and limb-darkened disk (LD) model.  The $V^2$ data are binned into 30 points for better visualization, and are shown as filled-dots with error bars. The blue curve shows the best-fit UD model with a reduced-$\chi^2$ of 0.37, obtained from the bootstrap process.  The best-fit UD diameter from bootstrap is 1.097$\pm$0.009 mas. The red line slightly below is the LD model with a  reduced-$\chi^2$ of 0.36. A power-law limb darkening \citep{Hestroffer1997} is assumed for the model. The best-fit LD diameter from bootstrap is 1.121$\pm$0.007 mas, corresponding to a radius of 1.625$\pm$0.011\rsun~at 13.49 pc.
}
\label{vis2_upsand}
\end{center}
\end{figure}

\begin{figure}[t]
\begin{center}
   \includegraphics[width=4in, angle=90]{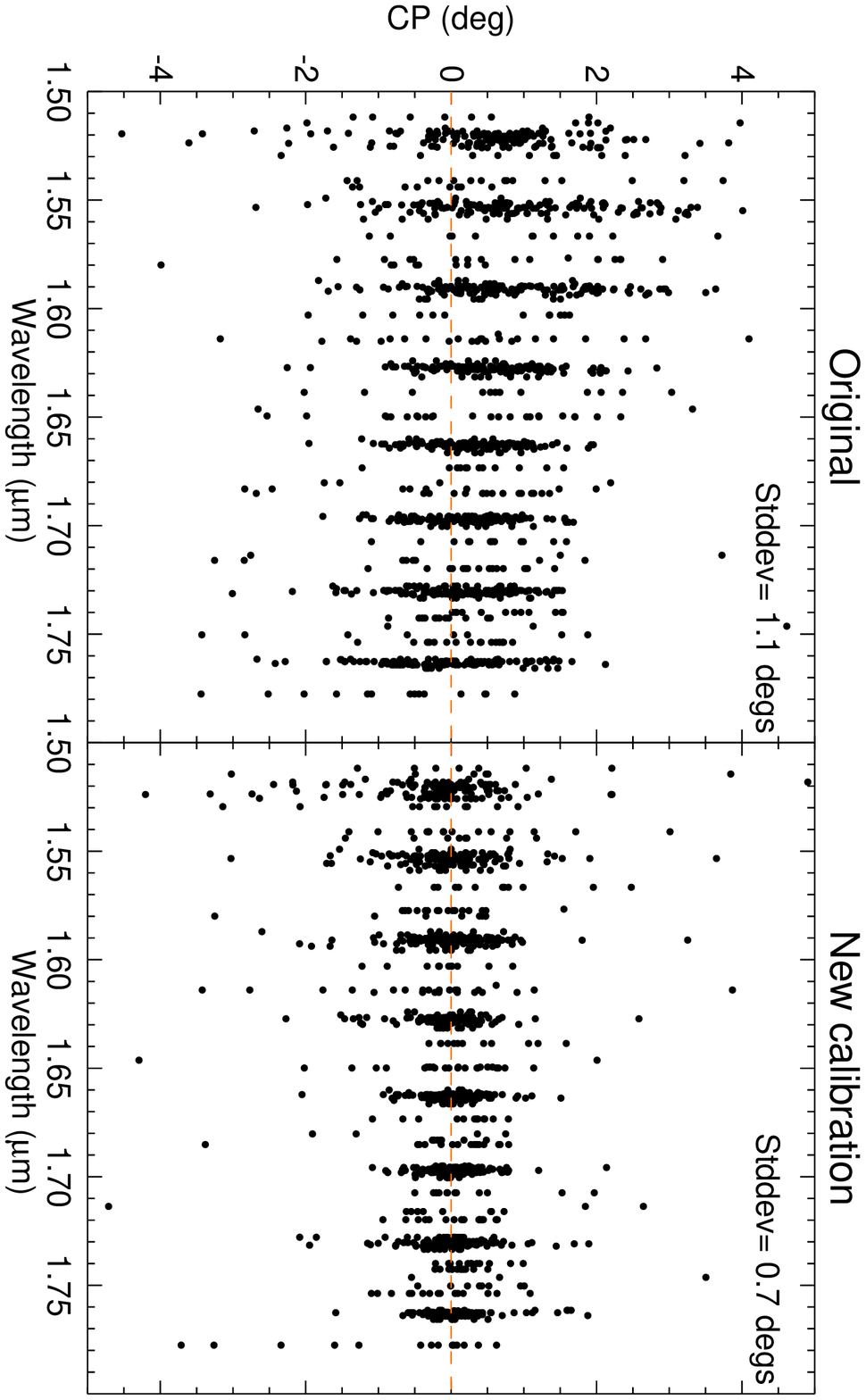}
\caption[The closure phases of $\upsilon$ And before and after the new calibration scheme]{The closure phases of $\upsilon$ And before (left panel) and after (right panel) the new calibration scheme. 
 Data from all observing nights are overplotted as a function of closure phase vs. wavelength. Due to the variation of wavelength calibration from night to night, the data are scattered around each of the eight wavelength channels of MIRC. The dashed lines represent the reference level for zero closure phases. The scatters of the data are shown in the top right corners. }
\label{cp_upsand}
\end{center}
\end{figure}



\section{Diameter and upper limits for $\upsilon$ And b}
\label{upsand}

\subsection{Refined diameter}
We first determine the angular diameter of $\upsilon$ And with a uniform disk (UD) model and a limb-darkened disk (LD) model. The squared-visibilities ($V^2$) of the data are calibrated in the regular way as described in \citet{Monnier2007sci}. UD diameters of the calibrators are listed in Table \ref{cals}. 
We apply a power law limb darkening, $I(\mu) = \mu^\alpha$, for the LD model \citep{Hestroffer1997}. Since the limb darkening coefficient $\alpha$ is similar for stars with the same spectral type, we adopt a fixed $\alpha$ and only vary the diameter in our LD fit.
The value of $\alpha$ is interpolated and converted from the square root law coefficients of \citet{VanHamme1993}. A value of $\alpha\simeq0.18$ is estimated for $\upsilon$ And, assuming [$Fe/H$]=0.15 \citep{Butler2006}.
We bootstrap\footnote{Bootstrapping is a technique that can provide robust simulations of the distribution of a data set. It is very useful for data sets with complicated or unknown distributions, and is widely used to derive estimates of standard errors and confidence intervals \citep{Press1992, Efron1993}. Since we use multiple nights of data with various of baselines and different systematics (within each night) for joint solutions, the distributions of our parameters of interest are unknown. Thus, bootstrapping is a suitable technique for our data. Bootstrapping requires the assumption that the data are independent and identically distributed. This assumption  holds for our data because we treat each night of data equally, each night's are independent and are acquired in the same way.} 
the $V^2$ data from different nights to simulate the statistics. Each bootstrapped data set is then fit with a UD and LD model separately with $\chi^2$ minimization. 
A number of 150 bootstrap iterations are carried out and the median of the 150 best-fit diameters are taken as the global best-fit, while the 1-$\sigma$ error is determined from the scatter. Figure \ref{vis2_upsand} shows the best-fit models of $\upsilon$ And, together with the binned data.
Our final best-fit UD of $\upsilon$ And is 1.097$\pm$0.009 mas, consistent with the FLUOR \citep{Coude2003} measurement of 1.098$\pm$0.007 mas (M{\'e}rand 2008, private communication), and the results of \citet{Baines2008}, 1.091$\pm$0.009 mas. The best-fit LD size is 1.121$\pm$0.007 mas, also consistent with the result of \citet{Baines2008}, and corresponding to a radius of 1.625$\pm$0.011\rsun~for the parallax of $74.12\pm0.19$ mas \citep{Leeuwen2007} (with error  propagated).

\subsection{Upper limits}
Since we do not have enough SNR to detect the planet from a single night of observation, we can take the advantage of the well known orbital parameters of $\upsilon$ And b and combine all observations  in Table \ref{obslog}  together to increase the total SNR for higher precision. 
 Due to the quickly varying closure phases caused by the Earth's rotation, as shown in Figure \ref{cp_figa}, we split the data into small chunks with an averaging time of less than 10min to avoid smearing the signal. 

To calibrate the large closure phase systematics described in last section, we apply the new calibration scheme by fitting a quadratic surface function to all  calibrators within each observing run, and subtracting the predicted zero closure phases from the raw values. For observing runs with inadequate calibrator coverages, we use the results from linear surface fit instead of quadratic fit. The total uncertainty of each data point is estimated by bootstrapping the data used in the quadratic or linear surface fit, and combining the additional uncertainty with the original values.
Figure \ref{cp_upsand} compares the results before and after applying the new calibration scheme for the 23 nights of data. 
As we can see in Figure \ref{cp_upsand}, the non-zero closure phases are calibrated out and the large scatters  are reduced in the newly calibrated data.

We then fit the new closure phase data with binary models to search for the closure phase signal from the planet. 
Because the closure phases are different for each channel, we determine  planet/star flux ratios for each of the 8 wavelength channels simultaneously at the same orbital position, and search for the best position with a joint $\chi^2$ minimization.  The orbital positions of the system are calculated by fixing the well known parameters from radial velocity observations, i.e., P, T$_0$, e, and $\omega$, and only varying the semi-major axis, $\Omega$, and inclination. The method and the corresponding codes are validated with data of the well-known binary $\iota$ Peg from \citet{Monnier2007}.

\begin{figure}[t]
\begin{center}
\includegraphics[width=5in, angle=90]{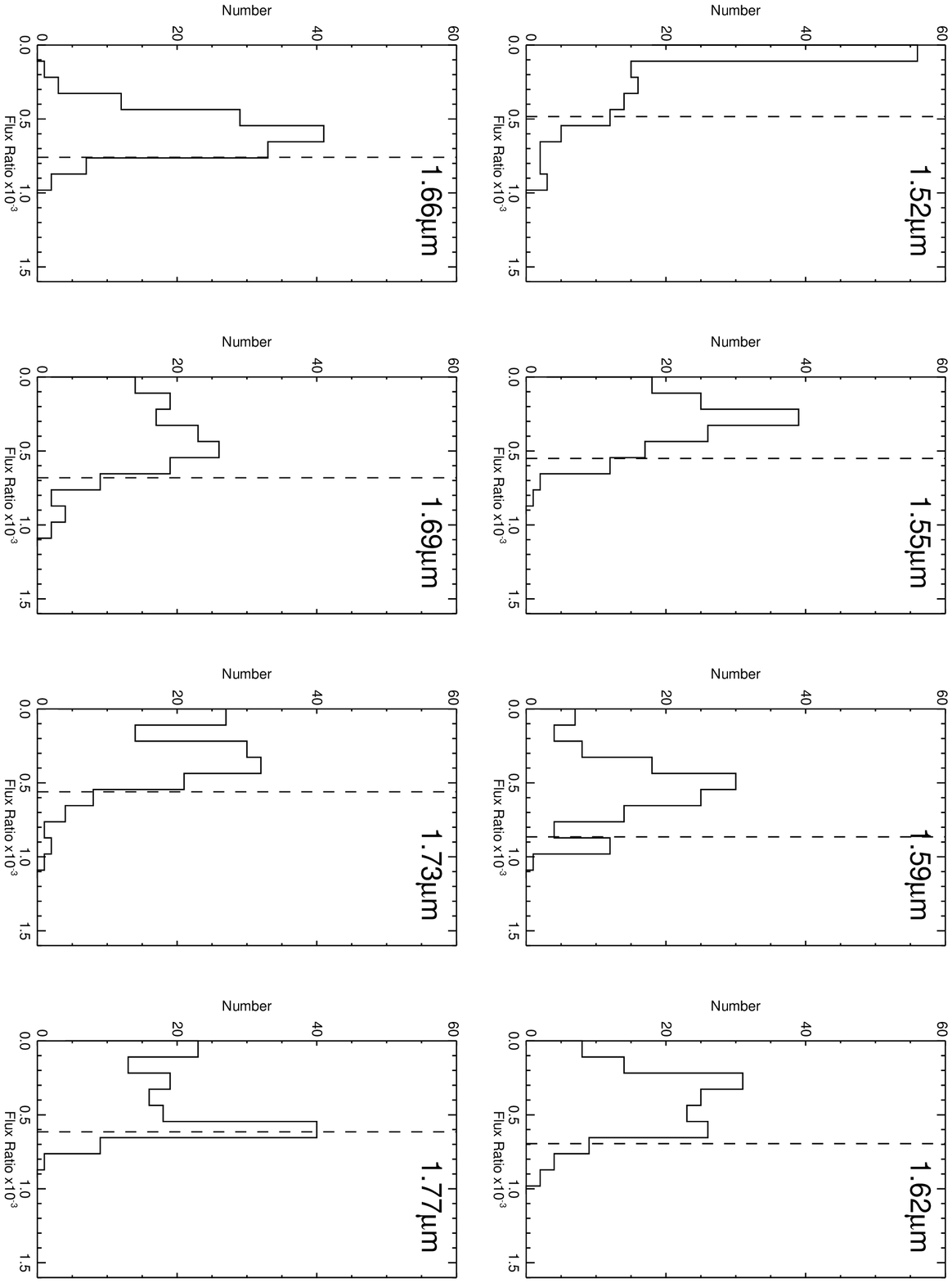}
\caption{Simulated distribution of the best-fit planet/star flux ratios of $\upsilon$ And b in the $H$ band, generated from $\sim$150 bootstraps of different nights of data. 
The dashed lines indicate the 90\% confidence levels for the upper limits.
}
\label{histo}
\end{center}
\end{figure}

Although we have searched the parameter space extensively, we do not clearly detect $\upsilon$ And b in our fits, suggesting that our current signal-to-noise ratio is still inadequate.  We thus decided to  report our results in terms of an upper limit to the planet/star flux  ratio. 
To do this, we simulate the statistics of the best-fit planet/star flux ratios by bootstrapping different nights of data. This approach treats each night of data equally, ensuring the robustness of the bootstrap by preventing  data from certain nights dominating the statistics. Because noise is dominant in the data, we use a fine grid search  to ensure robust results. For each bootstrapped data set, we searched an extensive range of semi-major axis, $\Omega$, inclination, and the planet/star flux ratios at each of the 8 spectral channels. The set of parameters that yields the minimum $\chi^2$ is then chosen as the ``best-fit". 
A total of 150 bootstrap iterations are carried out, and the corresponding distributions of the best-fit planet/star flux ratios are shown in Figure \ref{histo}. 

\begin{figure}[t]
\begin{center}
\includegraphics[width=3.5in, angle=90]{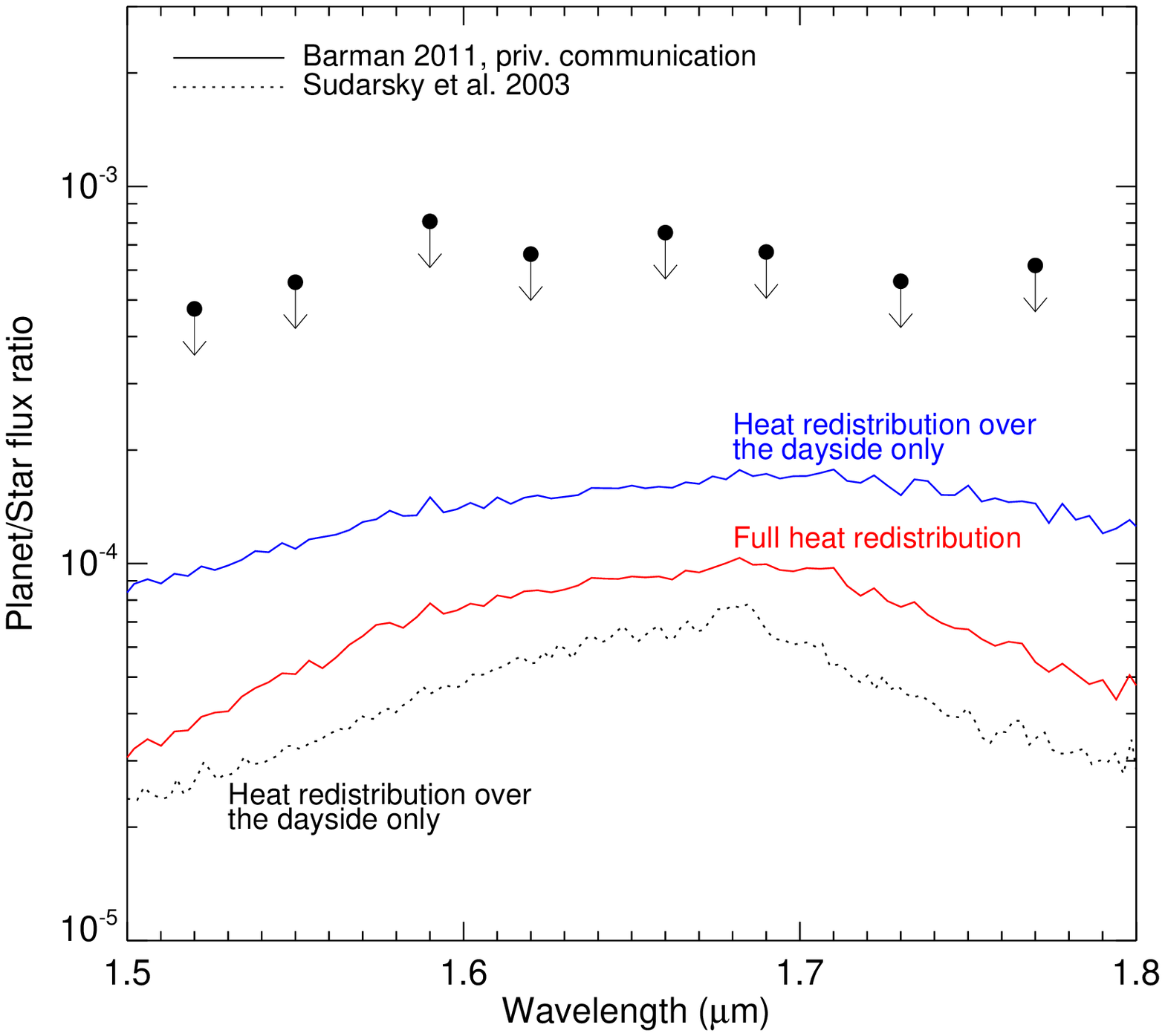}
\caption{Upper limits with 90\%  confidence levels for the planet/star flux ratios of $\upsilon$ And b in the $H$ band, using the newly calibrated CHARA/MIRC data from multiple nights. The average upper limit level is $6\times10^{-4}$. 
The best channel at 1.52$\mu m$ shows an upper limit flux ratio of 4.7$\times10^{-4}$. The solid lines show the latest model based on \citet{Barman2005} (Barman 2011, private communication), assuming a typical radius of 1.3 R$_j$ for the planet. The blue line shows the model with incident flux uniformly distributed over the dayside of the planet only, while the red line shows the model with full heat redistribution over the entire sphere.  The dotted line shows the model prediction from \citet{Sudarsky2003}, assuming a radius of 1 R$_j$ with heat redistribution over the dayside only.
}
\label{upper}
\end{center}
\end{figure}

 The dashed lines in Figure \ref{histo} indicate the upper limits of 90\% confidence level, i.e., with 90\% of chance these upper limits are higher than the actual planet/star flux ratios. 
Figure \ref{upper} shows the upper limits together as a ``spectrum", and compares it with planet atmospheric models based on \citet{Barman2005} (Barman 2011, private communication) and  \citet{Sudarsky2003}. Our upper limits are at the $6\times10^{-4}$ level on average. 
The first channel gives the best limit and reaches a level of 4.7$\times10^{-4}$, corresponding to a star/planet contrast ratio of 2.1$\times10^3$:1. This result stands as one of the highest contrast limits achieved by closure phase measurements to date.  

Our 90\% upper limits for the middle channels are about a factor of 5-8 from the predicted value of $\upsilon$ And b, suggesting that with further improvement in precision, we will be able to start constraining atmospheric models for hot Jupiters with interferometry.
In fact,  the precision of the new calibration for these data is not perfect due to their lack of wide calibrator coverages on the sky. In addition, although the new calibration can correct for the closure phase drifts caused by dispersion, the use of calibrators  from multiple nights makes the night-to-night variation hard to calibrate, leaving uncorrected systematic errors in the data. To reduce these effects and further improve our calibration precision, the new observing scheme using higher spectral resolution (using the grism of R=150 for MIRC in this case) and more calibrators is necessary. Benefitting from this experiment and analysis, better precision is expected in future observations.

\section{Conclusions and future prospects}
\label{conclusion}

We have simulated the closure phases for the best hot Jupiter candidate $\upsilon$ And b, and investigated the precision and stability of our measurements obtained with CHARA/MIRC.
Although our closure phase precisions can reach 0.3$^o$/1.5hrs for the middle wavelength channel of MIRC with good conditions, we have also encountered unexpected closure phase drifts as large as $8^o$ in other channels. The closure phase drifts are highly correlated with altitude and azimuth angles of the targets. The slopes of the drifts also vary across spectral channels, centering at the middle channels of MIRC. 
Because the drifts are correlated for all calibrators within an observing run, we are able to model the trend  to calibrate the drifts. This new calibration model, however, is highly dependent on the altitude and azimuth coverage of the calibrators and may not be accurate for sparse coverage. 
With a set of diagnostics, we find that the closure phase drifts are most likely caused by extra dispersion in the delaylines, which contaminates closure phases with non-closed triangles from other wavelengths, especially at the edges of the bandpass. Using higher spectral resolution can effectively reduce this effect.  We therefore advocate future observations of CHARA/MIRC to use the grism mode of R=150 for better calibration precision.

Taking advantage of the well known orbital parameters of $\upsilon$ And b, we have combined multiple nights of observations together for a joint solution of upper limits for the planet/star flux ratios across the $H$ band. 
Our best upper limit indicates a contrast ratio of about 2.1$\times10^3$:1 at 90\% confidence level, standing as one of the highest upper limits yet achieved by closure phase measurements. 
Future observations with reduced dispersion contaminations using the grism mode are expected to have better performance. 

Recently, photometric channels for real-time flux calibration have been implemented for MIRC \citep{Che2010}. The photometric channels have not only improved the visibility calibration, but also improved the data taking efficiency for MIRC by a factor of $\sim2$ by reducing the time spent on repositioning fibers of each MIRC beam. The CHAMP fringe tracker for MIRC \citep{Berger2008} has been commissioned in 2009 and is expected to be fully functional soon. CHAMP will help track and stabilize the fringes to increase their coherence time, allowing longer integration for MIRC and therefore can increase the SNR to the photon-limited regime and significantly improve the precision. 
In addition, CHARA is seeking an adaptive optics upgrade \citep{Ridgway2008} that could increase the throughput of MIRC by at least a factor of 2, which could greatly improve the sensitivity and data taking efficiency. 
With these implemented and upcoming improvements and the new observing scheme, we are expecting much better closure phase performance and are optimistic of achieving  the goal of detecting emission from hot Jupiters in the near future.

\acknowledgments

The CHARA Array is funded by the Georgia State University, by the National Science Foundation through grant
 AST-0908253, by the W. M. Keck Foundation, and by the NASA Exoplanet Science Institute.
Part of this research was carried out at the Jet Propulsion Laboratory, California Institute of Technology, under a contract with the National Aeronautics and Space Administration. This research was supported by the former Michelson Graduate Student Fellowship at the University of Michigan and the NASA Postdoctoral Program at the Jet Propulsion Laboratory (M. Z.).
J. D. M thanks the NSF grants AST-0352723, AST-0807577, and the NASA grant NNG04GI33G.
E. P. was formally supported by the Michelson Postdoctoral Fellowship and
is currently supported by a Scottish Universities Physics Association
(SUPA) advanced fellowship. STR acknowledges partial support from NASA grant NNH09AK731.
This research has made use of the SIMBAD database, operated at CDS, 
  Strasbourg, France, the Exoplanets Encyclopedia maintained by Jean Schneider at Paris Observatory, and the Exoplanet Orbit Database
at exoplanets.org.
  
\clearpage

\clearpage


\begin{deluxetable}{lccclc}
\tabletypesize{\scriptsize}
\tablecaption{Observation log}
\tablewidth{0pt}
\tablehead{
 \colhead{Observation} & \colhead{CHARA} &  Total integration\tablenotemark{a} & Spectral  & \colhead{Calibrators} & \colhead{Flux calibration} \\
 \colhead{Date} & \colhead{Telescope} & \colhead{time (min) }& Resolution &\colhead{for closure phase} & \colhead{method} 
 }
\startdata
 UT 2006Oct09   & S2-E2-W1-W2  & 8.03& R=50& $29$ Peg, $\zeta$ Per & chopper \\
 UT 2006Oct11	   & S2-E2-W1-W2  & 6.25& R=50&$29$ Peg, $\zeta$ Per & chopper \\ 
UT 2006Oct16	   & S2-E2-W1-W2  & 8.48& R=50&$29$ Peg, $\zeta$ Per & chopper \\
UT 2007Jul02	   & S1-E1-W1-W2  & 7.14& R=50&$\gamma$ Lyr, $\upsilon$ Peg, $\sigma$ Cyg & chopper \\ 
 UT 2007Jul04    & S1-E1-W1-W2  & 13.38 & R=50&$\gamma$ Lyr, $\upsilon$ Peg, $\sigma$ Cyg & chopper \\
 UT 2007Jul08    & S1-E1-W1-W2  & 8.92 & R=50&$\gamma$ Lyr, $\upsilon$ Peg, $\sigma$ Cyg & chopper \\
 UT 2007Jul29    & S2-E2-W1-W2  & 3.08 &R=50& $\sigma$ Cyg & chopper \\
 UT 2007Jul30    & S2-E2-W1-W2  & 3.50 & R=50&$\sigma$ Cyg & chopper \\
 UT 2007Aug02  & S2-E2-W1-W2  & 10.71 &R=50&$\sigma$ Cyg, 7 And, 37 And & chopper \\
 UT 2007Aug03  & S2-E2-W1-W2  &  9.82 &R=50&$\sigma$ Cyg, 7 And, 37 And & chopper \\
 UT 2007Aug06  & S2-E2-W1-W2  & 26.77 & R=50&$\sigma$ Cyg, 7 And, 37 And & chopper \\
 UT 2007Aug12  & S1-E1-W1-W2  & 40.15 &R=50&$\sigma$ Cyg, 7 And, 37 And & chopper\\
 UT 2007Aug13  & S1-E1-W1-W2  & 44.17 & R=50&$\sigma$ Cyg, 7 And, 37 And & chopper\\
 UT 2007Nov14  & S2-E2-W1-W2  & 36.58 &R=50&$\zeta$ Per, $\sigma$ Cyg, $\upsilon$ Peg, $\gamma$ Tri, 10 Aur  & chopper\\   
 UT 2007Nov16  & S2-E2-W1-W2  & 49.07 & R=50&$\zeta$ Per, $\sigma$ Cyg, $\upsilon$ Peg, $\gamma$ Tri, 10 Aur  & chopper\\
 UT 2007Nov17  & S2-E2-W1-W2  & 52.64 & R=50&$\zeta$ Per, $\sigma$ Cyg, $\upsilon$ Peg, $\gamma$ Tri, 10 Aur  & chopper\\
 UT 2007Nov19  & S1-E1-W1-W2  & 29.03 & R=50&$\zeta$ Per, 10 Aur, 70 Leo, 30 Leo  & chopper\\ 
 UT 2007Nov20  & S1-E1-W1-W2  & 30.14& R=50&$\zeta$ Per, 10 Aur, 70 Leo, 30 Leo  & chopper\\
 UT 2007Nov22  & S1-E1-W1-W2  & 52.46& R=50&$\zeta$ Per, 10 Aur, 70 Leo, 30 Leo  & chopper\\
 UT 2009Oct22\tablenotemark{b}   & S2-E1-W1-W2  & 19.63 & R=50&37 And, 10 Aur, $\epsilon$ Cas & Xchannel\tablenotemark{c} \\
 UT 2009Oct23\tablenotemark{b}   & S2-E1-W1-W2  & 15.18 & R=50&37 And, 10 Aur, $\epsilon$ Cas & Xchannel \\
 UT 2010Aug13\tablenotemark{d}  & S1-E1-W1-W2  & 9.30& R=150&10 Aur, $\gamma$ Tri & Xchannel \\
 UT 2010Aug14\tablenotemark{d} & S1-E1-W1-W2  & 9.96 & R=150&37 And, 10 Aur, $\gamma$ Tri & Xchannel \\
\enddata
\tablenotetext{a}{Data from each night are split into chunks with less than 10min of averaging time to avoid smearing of the closure phases.}
\tablenotetext{b}{Data taken with linear polarizer}
\tablenotetext{c}{Xchannel = photometric channel}
\tablenotetext{d}{Data taken with R=150 grism. The new calibration scheme is not applied to these data due to higher spectral resolution. For consistency, the 32 spectral channels are averaged into 8 channels in the fits as the other R=50 data.  }
\label{obslog}
\end{deluxetable}


\begin{deluxetable}{lcc}
\tabletypesize{\footnotesize}
\tablecaption{Adopted Parameters of $\upsilon$ And b}
\tablewidth{0pt}
\tablehead{\colhead{Parameter} & \colhead{Value} & \colhead{Ref.} }
\startdata
V (mag)		&	4.09		&  a\\
H (mag)		&	2.957 	&  a \\
K (mag)		&	2.859 	&  a \\
\hline
Distance (pc) & 13.49   & b \\
Period (days)	&  4.617136  & c \\
$e$		& 0.013 & c \\
Semimajor axis (AU) & 0.0595  & c  \\
Semimajor axis (mas) & 4.410 & c  \\
T$_{p}$ (JD)\tablenotemark{A} & 2454425.02 & d  \\ 
$w$ (deg) &  51$^\circ$ & d \\
Inclination (deg) \tablenotemark{B} & $\sim58^\circ$ & e \\
Stellar Diam. (mas) & 1.121 $\pm$ 0.007 & f \\
Stellar Diam. (\rsun) & 1.625 $\pm$ 0.011  & f \\
\enddata
\tablenotetext{~}{References: a. SIMBAD; b. \citet{Leeuwen2007}; c. \citet{Wright2009}; d. \citet{Butler2006}; e. \citet{Simpson2010}; f. {Limb-darkened} diameter from this work, see \S\ref{upsand} and Figure \ref{vis2_upsand}.}
\tablenotetext{A}{ Time of periastron passage}
\tablenotetext{B}{ Inclination of stellar rotation axis. Note the orbital inclination equals to this value only when the planet is coplanar with the star. }
\label{tab2}
\end{deluxetable}


\begin{deluxetable}{llcc}
\tabletypesize{\footnotesize}
\tablecaption{Results of calibration tests}
\tablewidth{0pt}
\tablehead{\colhead{Date} & \colhead{Observing} & \colhead{Magnitude of slope drift }\\
 & \colhead{Method}  & \colhead{(slope$_{ch8}$ - slope$_{ch1}$)}
} 
\startdata
2008 Aug 	22-28 \& Sep 02	& Original					& 0.37 $\pm$ 0.19 \\
2009 Nov 02 \& 04 	& Polarizer				& 0.13 $\pm$ 0.13 \\
2009 Dec 02, 03, 04 & Polarizer + 40$\mu m$ slit	& 0.18 $\pm$ 0.07 \\
2010 Aug 13 \& 14 	& Grism R=150				& 0.006 $\pm$ 0.067 \\
\enddata
\label{tab3}
\end{deluxetable}


\begin{deluxetable}{lcl}
\tabletypesize{\scriptsize}
\tablecaption{Calibrator diameters}
\tablewidth{0pt}
\tablehead{
\colhead{Calibrator} & \colhead{UD diameter (mas)} & \colhead{Reference} 
}
\startdata
29 Peg & 1.017 $\pm$0.027 & M{\'e}rand 2008, private communication \\
$\zeta$ Per   & 0.67 $\pm$ 0.03 & getCal\tablenotemark{a} \\
$\upsilon$ Peg & 1.01$\pm$ 0.04& \citet{Blackwell1994} \\
$\gamma$ Lyr &  0.742 $\pm$0.097& \citet{Leggett1986} \\
$\sigma$ Cyg & 0.542 $\pm$ 0.021 & M{\'e}rand 2008, private communication\\
7 And & 0.654 $\pm$ 0.029& Based on \citet{Kervella2008}\\
37 And&0.676 $\pm$ 0.034 & Based on \citet{Kervella2008} \\
$\gamma$ Tri & 0.522 $\pm$ 0.033 & \citet{vanBelle2008} \\
10 Aur &0.374 $\pm$ 0.079 & \citet{vanBelle2008}  \\
$\theta$ Leo & 0.677 $\pm$ 0.020 & Based on \citet{Kervella2008} \\ 
$\eta$ Leo & 0.710 $\pm$ 0.033 & Based on \citet{Kervella2008}\\
$\epsilon$ Cas & 0.375 $\pm$ 0.023 & Based on \citet{Barnes1978} \\
\enddata
\tablenotetext{a}{http://mscweb.ipac.caltech.edu/gcWeb/gcWeb.jsp}
\label{cals}
\end{deluxetable}



\clearpage
\bibliographystyle{apj}  
\bibliography{ms}  

\end{document}